\documentclass{article} 
\usepackage{nips13submit_e,times}
\usepackage{hyperref}
\usepackage{url}
\usepackage{graphicx}
\usepackage{algorithm}
\usepackage{algpseudocode}
\usepackage{amsfonts}
\usepackage{stmaryrd}
\algnewcommand\And{\textbf{ and }}
\algnewcommand\Or{\textbf{ or }}

\usepackage{tikz}
\usetikzlibrary{arrows,calc,shapes,decorations,decorations.pathreplacing}
\usepackage{xcolor}
\usepackage{etoolbox}
\newtoggle{quickdecim}

\title{Enigma: Decentralized Computation Platform with Guaranteed Privacy}

\author{
Guy Zyskind\\
\And
Oz Nathan\\
\And
Alex 'Sandy' Pentland\thanks{\texttt{guyz@mit.edu, oznathan@gmail.com, pentland@mit.edu}}\\
}

\nipsfinalcopy 

\begin{document}

\maketitle

\begin{abstract}

A peer-to-peer network, enabling different parties to jointly store and run computations on data while keeping the data completely private. Enigma's computational model is based on a highly optimized version of secure multi-party computation, guaranteed by a verifiable secret-sharing scheme. For storage, we use a modified distributed hashtable for holding secret-shared data. An external blockchain is utilized as the controller of the network, manages access control, identities and serves as a tamper-proof log of events. Security deposits and fees incentivize operation, correctness and fairness of the system. Similar to Bitcoin, Enigma removes the need for a trusted third party, enabling autonomous control of personal data. For the first time, users are able to share their data with cryptographic guarantees regarding their privacy.

\end{abstract}

\section{Motivation}

Since early human history, centralization has been a major competitive advantage. Societies with centralized governance were able to develop more advanced technology, accumulate more resources and increase their population faster \cite{intro1}. As societies evolved, the negative effects of centralization of power were revealed: corruption, inequality, preservation of the status quo and abuse of power. As it turns out, some ‘separation of powers’ \cite{intro2} is necessary. In modern times, we strive to find a balance between the models, maximizing output and efficiency with centralized control, guarded by checks and balances of decentralized governance. 

The original narrative of the web is one of radical decentralization and freedom\cite{intro3}. During the last decade, the web's incredible growth was coupled with increased centralization. Few large companies now own important junctures of the web, and consequently a lot of the data created on the web. The lack of transparency and control over these organizations reveals the negative aspects of centralization once again: manipulation \cite{intro4}, surveillance \cite{intro5}, and frequent data breaches \cite{intro6}.

Bitcoin \cite{nakamoto} and other blockchains \cite{crypto15} (e.g., Ethereum) promise a new future. Internet applications can now be built with a decentralized architecture, where no single party has absolute power and control. The public nature of the blockchain guarantees transparency over how applications work and leaves an irrefutable record of activities, providing strong incentives for honest behavior. Bitcoin the currency was the first such application, initiating a new paradigm to the web.

The intense verification and public nature of the blockchain limits potential use cases, however. Modern applications use huge amounts of data, and run extensive analysis on that data. This restriction means that only ‘fiduciary code’ can run on the blockchain \cite{intro7}. The problem is, much of the most sensitive parts of modern applications require heavy processing on private data. In their current design, blockchains cannot handle privacy at all. Furthermore, they are not well-suited for heavy computations. Their public nature means private data would flow through every full node on the blockchain, fully exposed. 
  
There is a strange contradiction in this setup. The most sensitive, private data can only be stored and processed in the centralized, less transparent and insecure model. We have seen this paradigm lead to catastrophic data leaks and the systematic lack of privacy we are currently forced to accept in our online lives.

\section{Enigma}

Enigma is a decentralized computation platform with guaranteed privacy. Our goal is to enable developers to build 'privacy by design', end-to-end decentralized applications, without a trusted third party. 

\textbf{Enigma is private}. Using \textit{secure multi-party computation} (\textit{sMPC} or \textit{MPC}), data queries are computed in a distributed way, without a trusted third party. Data is split between different nodes, and they compute functions together without leaking information to other nodes. Specifically, no single party ever has access to data in its entirety; instead, every party has a meaningless (i.e., seemingly random) piece of it.

\textbf{Enigma is scalable}. Unlike blockchains, computations and data storage are not replicated by every node in the network. Only a small subset perform each computation over different parts of the data. The decreased redundancy in storage and computations enables more demanding computations.

The key new utility Enigma brings to the table is the ability to run computations on data, without having access to the raw data itself. For example, a group of people can provide access to their salary, and together compute the average wage of the group. Each participant learns their relative position in the group, but learns nothing about other members' salaries. It should be made clear that this is only a motivating example. In practice, any program can be securely evaluated while maintaining the inputs a secret. 


Today, sharing data is an irreversible process; once it is sent, there is no way to take it back or limit how it is used. Allowing access to data for secure computations is reversible and controllable, since no one but the original data owner(s) ever see the raw data. This presents a fundamental change in current approaches to data analysis.

\section{Design overview}

Enigma is designed to connect to an existing blockchain and off-load private and intensive computations to an off-chain network. All transactions are facilitated by the blockchain, which enforces access-control based on digital signatures and programmable permissions.

Code is executed both on the blockchain (public parts) and on Enigma (private or computationally intensive parts). Enigma's execution ensures both \textit{privacy} and \textit{correctness}, whereas a blockchain alone can only ensure the latter. Proofs of correct execution are stored on the blockchain and can be audited. We supply a scripting language for designing end-to-end decentralized applications using \textit{private contracts}, which are a more powerful variation of \textit{smart contracts} that can handle private information (i.e., their state is not strictly public).

\begin{figure}[htbp]
\centering
\tikzstyle{sensor}=[draw, text width=5em, text centered, minimum height=2.5em]
\tikzstyle{naveqs} = [sensor, text width=6em, minimum height=12em, rounded corners]

\begin{center}
\begin{tikzpicture}
  \node at (6,2.7) (x0) [naveqs] {Blockchain};
  \node at (9,2.7) (x1) [naveqs] {Enigma};
  \node at (2,1) {$f : X \rightarrow Y$};
  \begin{scope}[every node/.style={draw, anchor=text, rectangle split,
    rectangle split parts=7,minimum width=2cm}]
    \node (R) at (2,4){ \nodepart{two} \nodepart{three}
    \nodepart{four}$op_i$\nodepart{five}\nodepart{six}\nodepart{seven}};
  \end{scope}
  \draw[decorate,decoration={brace,mirror,raise=6pt,amplitude=5pt}, thick]
    (R.north west)--(R.three west) ;
  \draw[decorate,decoration={brace,raise=6pt,amplitude=10pt}, thick]
    (R.three east)--(R.south east); 
  \draw[->] ($(R.west)+(-20pt,25pt)$) to[out=-180,in=240] ++(0,1.5)
    to [out=40,in=120]node[above,midway]{public}(x0) ; 
  \draw[->] ($(R.east)+(20pt,-40pt)$) 
    to [out=-60,in=-120]node[below,midway]{private}(x1) ;
\end{tikzpicture}
\end{center}
\caption{Code execution model.}
\label{fig:code_exec}
\end{figure}
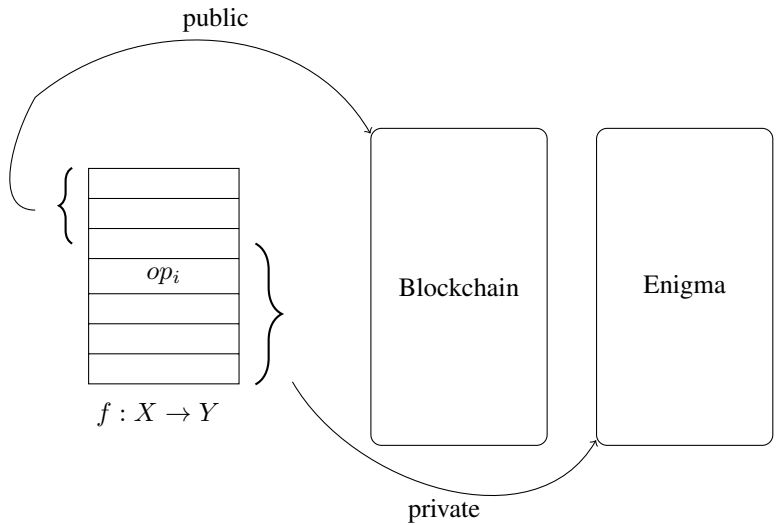

The scripting language is also turing-complete, but this is not as important as its scalability. Code execution in blockchains is decentralized but not distributed, so every node redundantly executes the same code and maintains the same public state. In Enigma, the computational work is efficiently distributed across the network. An interpreter breaks down the execution of a private contract, as is illustrated in Figure \ref{fig:code_exec}, resulting in improved run-time, while maintaining both privacy and verifiability.

The off-chain network solves the following issues that blockchain technology alone cannot handle:


\begin{enumerate}
	\item \textbf{Storage}. Blockchains are not general-purpose databases. Enigma has a decentralized off-chain \textit{distributed hash-table} (or \textit{DHT}) that is accessible through the blockchain, which stores references to the data but not the data themselves. Private data should be encrypted on the client-side before storage and access-control protocols are programmed into the blockchain. Enigma provides simple APIs for these tasks in the scripting language.
	
	\item \textbf{Privacy-enforcing computation}. Enigma's network can execute code without leaking the raw data to any of the nodes, while ensuring correct execution. This is key in replacing current centralized solutions and trusted overlay networks that process sensitive business logic in a way that negates the benefits of a blockchain. The computational model is described in detail in section \ref{sec:computation}.
	
	\item \textbf{Heavy processing}. Even when privacy is not a concern, the blockchain cannot scale to clearing many complex transactions. The same off-chain computational network is used to run heavy publicly verifiable computations that are broadcast through the blockchain.
	
\end{enumerate}

\section{Off-chain storage}
\label{sec:storage}

\begin{figure}[htbp]
\centering
\begin{center}
\def\mystrut{\vrule height 0.5cm depth 0.5cm width 0pt}
\begin{tikzpicture}
  \begin{scope}[every node/.style={draw, rectangle split,
    rectangle split parts=3,minimum width=2cm,minimum height=8cm}]
    \node (rect) at (2,4){\mystrut shares \nodepart{two}{\mystrut encrypted data} \nodepart{three}{\mystrut public data}
    };
  \end{scope}
\end{tikzpicture}
\end{center}
\caption{A node's local view of the off-chain data.}
\label{fig:node_view}
\end{figure}
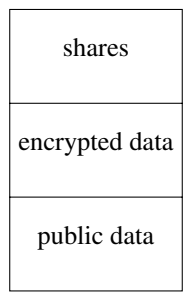

Off-chain nodes construct a distributed database. Each node has a distinct view of shares and encrypted data so that the computation process is guaranteed to be privacy-preserving and fault tolerant. It is also possible to store large public data (e.g., files) unencrypted and link them to the blockchain. Figure \ref{fig:node_view} illustrates the database view of a single node.

On a network level, the distributed storage is based on a modified Kademlia DHT protocol \cite{kademlia} with added persistence and secure point-to-point channels, simulated using a broadcast channel and public-key encryption. This protocol assists in distributing the shares in an efficient manner. When storing shares, the original Kademlia distance metric is modified to take into account the preferential probability of a node.


\section{Privacy-enforcing computation}
\label{sec:computation}

In this section, we describe Enigma's computational model. We begin with a brief introduction to publicly verifiable secure MPC based on state-of-the-art advances in cryptography. Then, we describe a series of performance improvements to secure MPC that makes the technology practical even when the network is large: hierarchical secure MPC, network reduction and adaptable circuits.

To use Enigma, developers write high-level code, where public parts are executed on the blockchain and private parts are run off-chain, on Enigma's platform. We call these \textit{private contracts}, since they are smart contracts that can handle private information.


\subsection{Overview of secure multi-party computation}

\subsubsection{Privacy (passive adversaries)}

Yao introduced the first solution to secure two-party computation protocols in 1982 \cite{yao82}. In the same paper, Yao suggested the popular \textit{millionaire problem}, describing two millionaires interested in knowing which one of them is richer, without revealing their actual net worth. In the decades since, the two-party problem has been generalized to MPC, which refers to the $n$-party case. For general-purpose MPC, in which every protocol could be composed from a circuit of elementary MPC gates, two major approaches have been developed over the years: Yao's garbaled (boolean) circuits \cite{fairplaymp} and MPC based on secret sharing. The latter has been more commonly used in production systems (e.g., \cite{sharemind} and \cite{viff}) and is our focus as well.

A threshold cryptosystem is defined by $(t+1,n)-threshold$, where $n$ is the number of parties and $t+1$ is the minimal number of parties required to decrypt a secret encrypted with threshold encryption. Secret sharing is an example of a threshold cryptosystem, where a secret $s$ is divided among $n$, s.t. at least $t+1$ are required to reconstruct $s$. Any subset of $t$ parties cannot learn anything about the secret. A \textit{linear secret-sharing scheme} (or \textit{LSSS}) partitions a secret to shares such that the shares are a linear combination of the secret. \textit{Shamir's secret sharing} (or \textit{SSS}) is an example of a LSSS, which uses polynomial interpolation and is secure under a finite field $\mathbb{F}_p$ \cite{shamir}. Specifically, to share a secret $s$, we select a random $t$ degree polynomial $q(x)$ --

\begin{eqnarray}
	q(x) = a_0 + a_1 x + \cdots + a_{t} x^{t},\\
	a_0 = s, a_i \sim U(0,p-1).
\end{eqnarray}

The shares are then given by 
\begin{equation}
	\forall i \in \{ 1, \cdots, n \}: [s]_{p_i} = q(i).
\end{equation}

Then, given any $t+1$ shares, $q(x)$ could be trivially reconstructed using Lagrange interpolation and the secret $s$ recovered using $s = q(0)$. Since SSS is linear, it is also additively homomorphic, so addition and multiplication by a scalar operations could be performed directly on the shares without interaction. Formally --

\begin{eqnarray}
	c \times s = reconstruct( \{ c[s]_{p_i} \}_{i \in n}^{t+1}),\\
	s_1 + s_2 = reconstruct( \{ [s_1]_{p_i} + [s_2]_{p_i} \}_{i \in n}^{t+1}).
\end{eqnarray}

Multiplication of two secrets $s_1$ and $s_2$ is somewhat more involved. If each party would attempt to locally compute the product of two secrets, they would collectively obtain a polynomial of degree $2t$, requiring a polynomial reduction step ($2t \rightarrow t$). For an information theoretic setting, this result adds an honest majority constraint (i.e., $t < \frac{n}{2}$) on privacy and correctness. If we bound the adversary's computational power, both properties are assured for any number of corrupted parties, but fairness and deciding on an output still requires an honest majority \cite{GBW88}.

As to performance, a re-sharing step is required in the degree reduction step, implying all parties must interact with all other parties ($O(n^2)$ communications). This makes MPC impractical for anything larger than a small constant number of parties $n$. While optimized solutions exist for improving the amortized complexity, they are based on assumptions that restrict functionality in practice. Conversely, we describe a generic solution to this problem for any functionality in Section \ref{sec:hmpc}, which makes secure MPC feasible for arbitrarily large networks.

Note that with secure addition and multiplication protocols, we can construct a circuit for any arithmetic function. For turing-completeness, we need to handle control flow as well. For conditional statements involving secret values, this means evaluating both branches and for dynamic loops we add randomness to the execution. Our general-purpose MPC interpreter is based on these core concepts and other optimizations presented throughout the paper.


\subsubsection{Correctness (malicious adversaries)}
\label{sec:comp_correct}

So far we have discussed the \textit{privacy} property. \textit{Liveness}, namely -- that computations will terminate and the system will make progress, is also implied given an honest majority, since it is all that is needed for reconstruction of intermediate and output values. However, in the current framework there are no guarantees about the \textit{correctness} of the output; party $p_i$ could send an invalid result throughout the computation process which may invalidate the output. While BGW \cite{GBW88} presented an information-theoretic solution to verifiable MPC, its practical complexity could be as bad as $O(n^8)$, given a naive implementation \cite{TODO}.

Therefore, our goal is to design an MPC framework that is secure against malicious adversaries but has the same complexity of the semi-honest setting ($O(n^2)$). Later, we would further optimize this as well.

Very recently, Baum et al. developed a publicly auditable secure MPC system that ensures correctness, even when all computing nodes are covertly malicious, or all but a single node are actively malicious \cite{baum14}. Their state-of-the-art results are based on a variation of SPDZ (pronounced \textit{speedz}) \cite{spdz} and depend on a public append-only bulletin board, which stores the trail of each computation. This allows any auditing party to check the output is correct by comparing it to the public ledger's trail of proofs. Our system uses the blockchain as the bulletin board, thus our overall security is reduced to that of the hosting blockchain.

\textbf{SPDZ}. A protocol secure against malicious adversaries (with dishonest majority), providing correctness guarantees for MPC. In essence, the protocol is comprised of an expensive offline (pre-processing) step that uses \textit{somewhat homomorphic encryption} (or \textit{SHE}) to generate shared randomness. Then, in the online stage, the computation is similar to the passive case and there is no expensive public-key cryptography involved. In the online stage, every share is represented by the additive share and its MAC, as follows:

\begin{eqnarray}
	\langle s \rangle_{p_i} = ([s]_{p_i}, [\gamma(s)]_{p_i}),\ s.t.\ 
	\gamma(s) = \alpha s,
\end{eqnarray}

where $\alpha$ is a fixed secret shared MAC key and $\langle \bullet \rangle$ denotes the modified secret sharing scheme which is also additively homomorphic. $\langle \bullet \rangle$-sharing works without opening the shares of the global MAC key $\alpha$, so it can be reused.

As before, multiplication is more involved. Multiplication consumes $\{ \langle a \rangle, \langle b \rangle, \langle c \rangle \}$ triplets, s.t. $c = ab$, that are generated in the pre-processing step (many such triplets are generated). Then, given two secrets $s_1$ and $s_2$, that are shared using $\langle \bullet \rangle$-sharing, secret-sharing the product $s = s_1 s_2$ is achieved by consuming a triplet as follows --

\begin{eqnarray}
	\langle s \rangle = \langle c \rangle + \epsilon \langle b \rangle + \delta \langle a \rangle + \epsilon \delta,\\
	\epsilon = \langle s_1 \rangle - \langle a \rangle,\ \delta = \langle s_2 \rangle - \langle b \rangle.
\end{eqnarray}

As mentioned, generating the triplets is an expensive process based on SHE. The full protocol including security proofs is found in \cite{baum14}. Verification is achieved by solving --

\begin{equation}
	\gamma - \alpha s = 0,
\end{equation}

where $s$ is the secret that, without loss of generality, can be the reconstructed result of any secure computation. Intuitively, this is just a comparison of the computation over the MAC against the computed result times the secret MAC key. The reason we are not performing actual comparison is so that $\alpha$ remains secret and can be reused.

We can now see that $\langle \bullet \rangle$-sharing has similar properties to SSS, namely -- it is additively homomorphic and requires a re-sharing round for multiplication ($O(n^2)$ communication complexity), but in addition -- it ensures correctness against up to $n-1$ active adversaries. The offline round is easily amortized over many computations and can be computed in parallel while other computations are running, so it does not significantly affect the overall efficiency.

\textbf{Publicly verifiable SPDZ}. In the publicly verifiable case, MACs and commitments are stored on the blockchain, therefore making the scheme secure even if all $n$ computing parties are malicious. We follow the representation of \cite{baum14}, which defines $\llbracket \bullet \rrbracket$-sharing, as --

\begin{eqnarray}
	\llbracket s \rrbracket = (\langle s \rangle, \langle r \rangle, \langle g^s h^r \rangle),
\end{eqnarray}

where $s$ is the secret, $r$ is a random value and $c=g^s h^r$ is the Pedersen commitment, with $g,h$ serving as generators. $\llbracket \bullet \rrbracket$-sharing preserves additive homomorphic properties, and with a slightly modified multiplication protocol we can re-use the same idea of generating triplets ($\{ \llbracket a \rrbracket, \llbracket b \rrbracket, \llbracket c \rrbracket \}$) offline.

A key observation here is that the nodes only need to compute over $\langle \bullet \rangle$-shared values and not over the commitments. These are stored on the blockchain and could later be addressed by any public validator that has the output. Even if a single node has broken its commitment, it would be evident to the auditor.







\subsection{Hierarchical secure MPC}
\label{sec:hmpc}

Information-theoretic results show that secure MPC protocols require each computing node to interact with all other nodes ($O(n^2)$ communication complexity) and a constant number of rounds. In the case of a LSSS, this computational complexity applies to every multiplication operation, whereas addition operations can be computed in parallel, without intercommunication. As previously mentioned, secure addition and multiplication protocols are sufficient to construct a general-purpose interpreter that securely evaluates any code \cite{GBW88}.

\begin{figure}[htbp]
\centering
\includegraphics[width=2.7in]{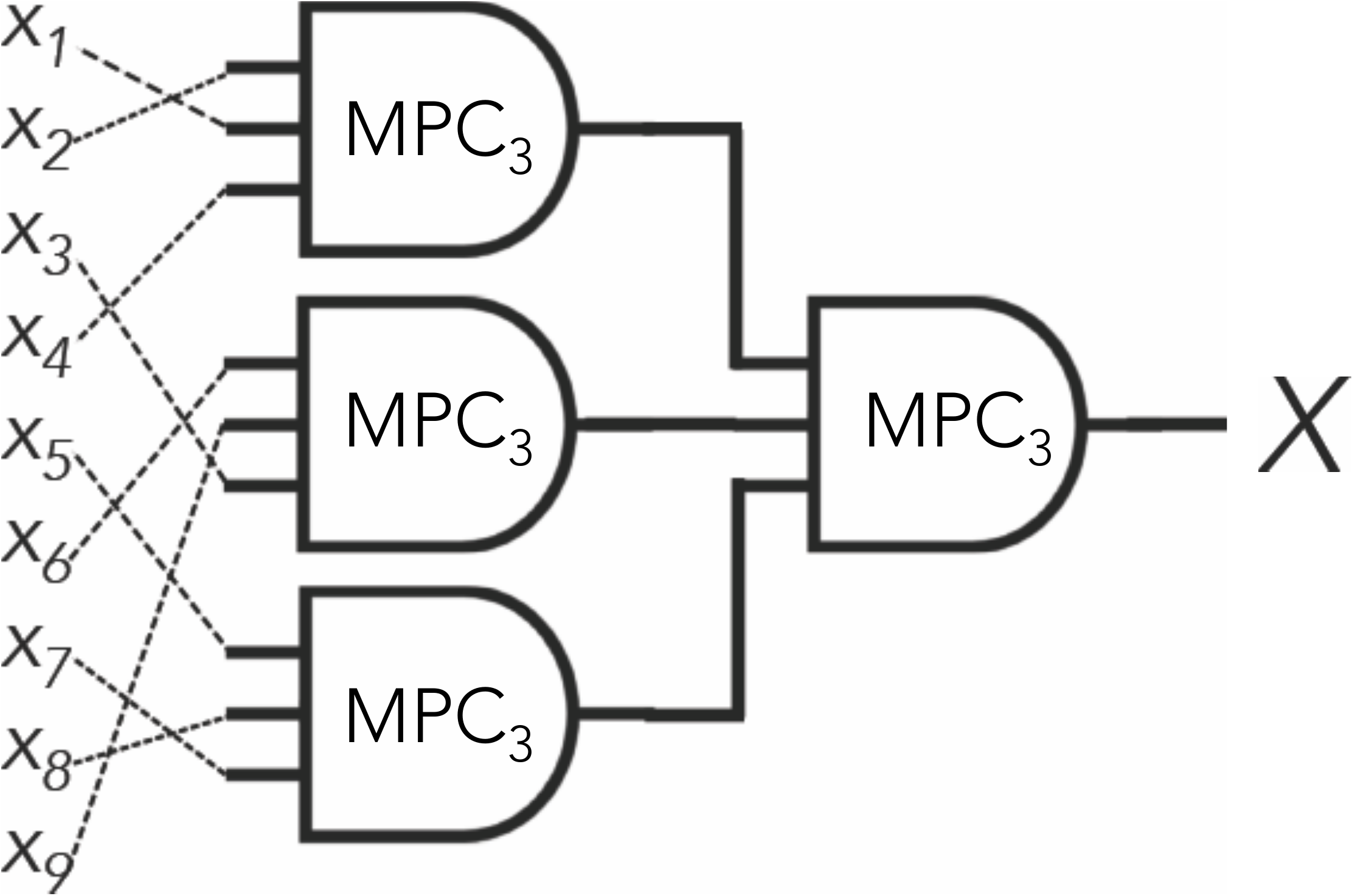}
\caption{Hierarchical Formula Builder.}
\label{fig:builder2}
\end{figure}

\begin{figure}[htbp]
\includegraphics[width=6in]{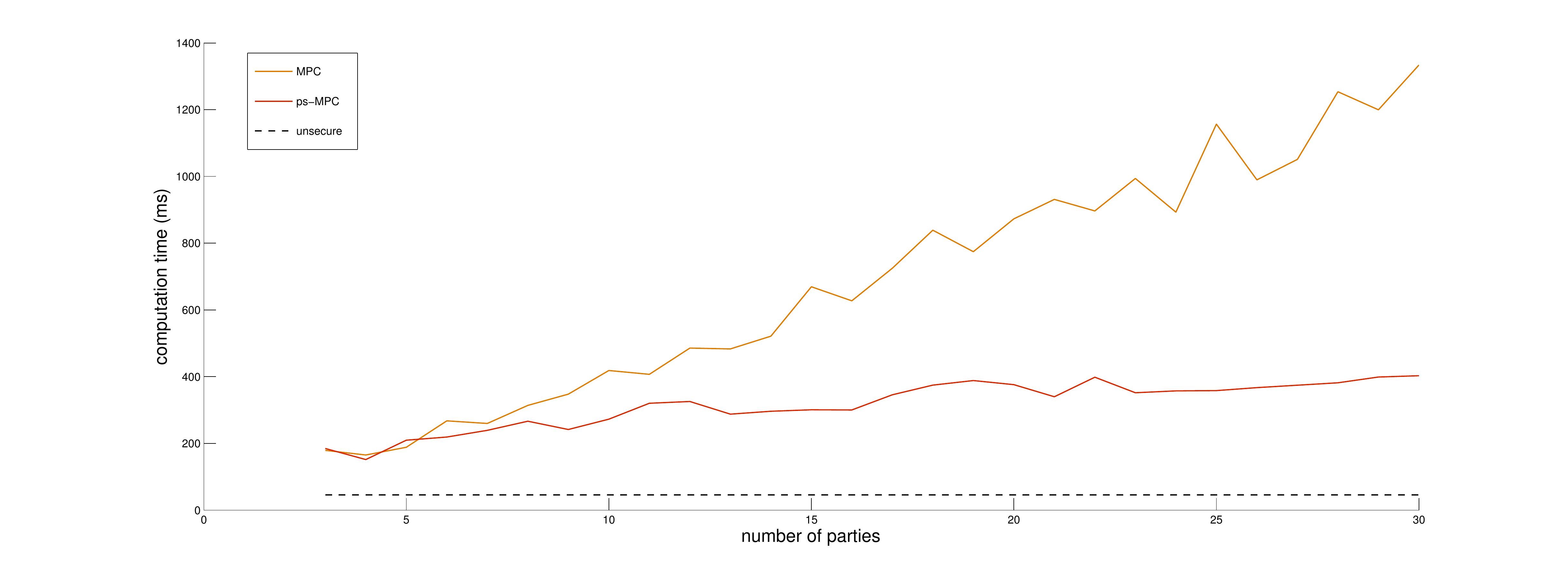}
\caption{Simulated performance comparison of our optimized secure MPC variant compared to classical MPC.}
\label{fig:mpc_sim2}
\end{figure}

Cohen et al \cite{cohen13} recently proposed a method of simulating an $n$-party secure protocol using a log-depth formula of constant-size MPC gates, as illustrated in Figure \ref{fig:builder2}. We extend their result to LSSS and are able to reduce the communication-complexity of multiplication from quadratic to linear, at the cost of increased computation complexity, which is parallelized. Figure \ref{fig:mpc_sim2} illustrates how vanilla MPC is limited by the number of parties, while our implementation scales up to arbitrarily large networks.


\subsection{Network reduction}

To maximize the computational power of the network, we introduce a network reduction technique, where a random subset of the entire network is selected to perform a computation. The random process preferentially selects nodes based on load-balancing requirements and accumulated reputation, as is measured by their publicly validated actions. This ensures that the network is fully utilized at any given point.


\subsection{Adaptable circuits}

Code evaluated in our system is guaranteed not to leak any information unless a dishonest majority colludes ($t \geq \frac{n}{2}$). This is true for the inputs, as well as any interim variables computed while the code is evaluated. An observant reader would notice that as a function is evaluated from inputs to outputs, the interim results generally become less descriptive and more aggregative.

For simple functions or functions involving very few inputs, this may not hold true, but since these functions are fast to compute - no additional steps are needed.

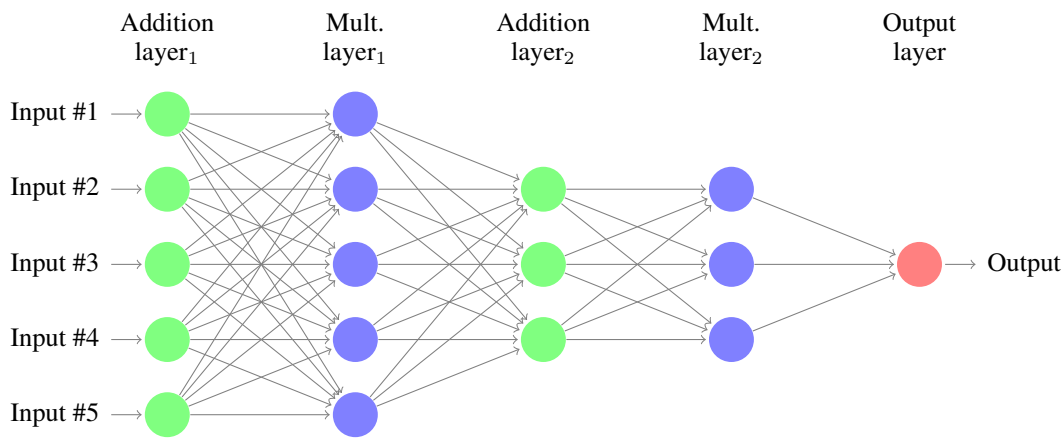
\begin{figure}[htbp]
\centering

\begin{center}
\def\layersep{2.5cm}

\begin{tikzpicture}[shorten >=1pt,->,draw=black!50, node distance=\layersep]
    \tikzstyle{every pin edge}=[<-,shorten <=1pt]
    \tikzstyle{neuron}=[circle,fill=black!25,minimum size=17pt,inner sep=0pt]
    \tikzstyle{input neuron}=[neuron, fill=green!50];
    \tikzstyle{output neuron}=[neuron, fill=red!50];
    \tikzstyle{hidden neuron}=[neuron, fill=blue!50];
    \tikzstyle{annot} = [text width=4em, text centered]

    \foreach \name / \y in {1,...,5}
        \node[input neuron, pin=left:Input \#\y] (I-\name) at (0,-\y) {};

    \foreach \name / \y in {1,...,5}
        \path
            node[hidden neuron] (H-\name) at (\layersep,-\y cm) {};

    \foreach \name / \y in {1,...,3}
        \path[yshift=-1cm]
            node[input neuron] (H1-\name) at (2*\layersep,-\y cm) {};
            
    \foreach \name / \y in {1,...,3}
        \path[yshift=-1cm]
            node[hidden neuron] (H2-\name) at (3*\layersep,-\y cm) {};
            
    \node[output neuron,pin={[pin edge={->}]right:Output}, right of=H2-2] (O) {};

    \foreach \source in {1,...,5}
        \foreach \dest in {1,...,5}
            \path (I-\source) edge (H-\dest);

    \foreach \source in {1,...,5}
        \foreach \dest in {1,...,3}
            \path (H-\source) edge (H1-\dest);
            
    \foreach \source in {1,...,3}
        \foreach \dest in {1,...,3}
            \path (H1-\source) edge (H2-\dest);
            
    \foreach \source in {1,...,3}
        \path (H2-\source) edge (O);

    \node[annot,above of=I-1, node distance=1cm] (il) {Addition layer$_1$};
    \node[annot,right of=il] (hl) {Mult. layer$_1$};
    \node[annot,right of=hl] (hl1) {Addition layer$_2$};
    \node[annot,right of=hl1] (hl2) {Mult. layer$_2$};
    \node[annot,right of=hl2] (ol) {Output layer};
\end{tikzpicture}
\end{center}

\caption{Feed forward flow of the secure code evaluation.}
\label{fig:ffnet}
\end{figure}

However, for computationally expensive functions, involving many lines of code and a large number of inputs, we can dynamically reduce the number computing nodes as we progress, instead of having a fixed $n$ for the entire function evaluation process. Specifically, we design a feed-forward network (Figure \ref{fig:ffnet}) that propagates results from inputs to outputs. The original code is reorganized so that we process addition gates on the inputs first, followed by processing multiplication gates. The interim results are then secret-shared with $\frac{N}{c}$ nodes, and the process is repeated recursively.

\subsection{Scripting}

As previously mentioned, end-to-end decentralized apps are developed using \textit{private contracts}, which are further partitioned to on-chain and off-chain execution. Off-chain code returns results privately, while sending correctness proofs to the blockchain. For simplicity, the scripting language is similar in syntax to well-known programming languages.

There are two major additions to the scripting language that require more detail.

\subsection{Private data types}

Developers should use the \textit{private} keyword to specify private objects. This automatically ensures that any computation involving those objects remains secure and private. When working with private objects, the data themselves are not locally available, but rather a reference of them.

\subsection{Data access}

There are three distinct decentralized databases living in the system, each accessible through a global singleton dictionary. Specifically --

\begin{enumerate}
	\item \textbf{Public ledger}. The blockchain's public ledger can be accessed and manipulated using $L$. For example, $L[k] \gets 1$ would update key $k$ for all nodes. Since the ledger is completely public and append-only, the entire history is stored as well and (read-only) accessible using $L.get(k,t)$.
	
	\item \textbf{DHT}. Off-chain data are stored on the DHT and accessible in the same way the public ledger is. By default, data are encrypted locally before transmission and only the signing entity can request the data back. Otherwise, using $DHT.set(k,v,p)$, where $k$ is the key, $v$ is the value and $p$ is a predicate, namely -- $p: X \rightarrow \{ 0, 1 \}$, sets $v$ to be accessible through $k$ if and only if $p$ is satisfied. We supply several built-in predicates in the language such as limiting access to a list of public keys. If encryption is turned off, the default predicate is $\forall x\ p(x) = 1$, so the data are public but distributed off-chain.
	
	\item \textbf{MPC}. Syntactically, using MPC is equivalent to DHT, but the underlying process differs. In particular, executing $MPC.set(k,v,p)$ secret shares $v$. The shares are distributed to potential computing parties that store their shares in their local view. Now $p$ can be used to specify who can reference the data for computation using $v_{ref} \gets MPC[k]$, without revealing $v$. By default, only the original dealer can ask for the raw data back by running $v \gets MPC.declassify(k)$, which similar to the sharing process, collects shares from the various parties and reconstructs the secret value locally. In addition, any other entities belonging to the same shared identity can reference the data for computation. For details about shared identities see section \ref{sec:identities}.
\end{enumerate}

Note that for simplicity, we addressed all keys in $L$, $DHT$ and $MPC$ dictionaries as using a single namespace, whereas in practice finer granularity is available, so that they can be segmented to databases, tables, and finer hierarchies. 





\section{Blockchain interoperability}

In this section we show how Enigma interoperates with a blockchain. Specifically, we detail how complex identities are formed using digital signatures, which are automatically compatible with blockchains. We then continue to describe in detail the core protocols linking Enigma's off-chain storage and computation to a blockchain.


\subsection{Identity management}
\label{sec:identities}

A recent survey paper divided blockchain-inspired technologies into two: fully decentralized permission-less ledgers (e.g., Bitcoin, Ethereum) and semi-centralized permissioned ledgers (e.g., Ripple) \cite{consensusaas}. In the paper, the author argues that there is an inherent trade-off between having a pseudo-anonymous system, where no one is trusted and all information must remain public, and having a somewhat centralized system with trusted nodes that can verify true underlying identities. With an off-chain technology linked to a blockchain, this trade-off can be avoided while the network remains fully decentralized.

For this to work, we define an extended version of identities, one that captures \textit{shared identities} across multiple entities and their semantic meaning. Formally, the pseudo-anonymous portion of a shared identity is a $(2n+1)$-tuple -- 

\begin{equation}
	SharedIdentity_{P} = (addr_P, pk^{(p_1)}_{sig}, pk^{(p_2)}_{sig}, \cdots, pk^{(p_n)}_{sig})
\end{equation}

where $n$ denotes the number of parties. It should be clear that for $n=1$ we revert to the special pseudo-identity case.

To complete our definition of shared identities, we incorporate the idea of meta-data. Meta-data encapsulates the underlying semantic meaning of an identity. Primarily, these include public access-control rules defined by the same predicates mentioned earlier, which the network uses to moderate access-control, along with any other public or private data that is relevant.

For example, Alice may want to share with Bob her height, but not her weight. Alternatively, she may not even want to tell Bob her exact height, but will allow him to use her height in aggregate computations. In this case, Alice and Bob can establish a shared identity for this purpose. Alice invokes a private contract that shares her height using $MPC['alice\_height'] = alice\_height$, which Bob can reference for computations, without accessing Alice's height value directly. 

The default MPC predicate establishes that Alice's pseudonym is the owner of the shared information and that Bob has restricted access to it. The predicate, shared identity's list of addresses and a reference to the data are stored on the blockchain and collectively define the public meta-data, or in other words - information related to the identity that is not sensitive but should be used to publicly verify access rights. Any additional meta-data that is private, or in other words that only Alice, Bob and perhaps several others should have access to could be securely stored off-chain using the DHT.

It should now be clear how our system solves the need for trusted nodes. As always, public transactions are validated through the blockchain. With shared identities and predicates governing access-control stored on the ledger, the blockchain can moderate access to any off-chain resources. For anything else involving private meta-data, the off-chain network can act as a trustless privacy-preserving verifier.

\subsection{Link protocols}

We now discuss the core protocols linking the blockchain to off-chain resources. Specifically, we elaborate on how identities are formed and stored on the ledger; and how off-chain storage (DHT) and computation (MPC) requests are routed through the blockchain, conditional on satisfying predicates.

\subsubsection{Access control}

Protocol \ref{alg:shared_identity} describes the process of creating a shared identity and Protocol \ref{alg:check_policy} implements the publicly-verifiable contract for satisfying predicates.

\begin{algorithm}
\caption{Generating a shared identity}\label{alg:shared_identity}
\begin{algorithmic}
\Require $P = \{ p_i \}_{i=1}^N$ parties, $A = \{ POLICY_{p_i} \}_{i=1}^N$
\Ensure Ledger $L$ stores reference to the shared identity.
\State $addr_P = 0$
\State $ACL = \emptyset$
\For{$p_i \in P$}
	\State $(pk^{(p_i)}_{sig},sk^{(p_i)}_{sig})\gets \mathcal{G}_{sig}()$
	\State $addr_P = addr_P \oplus pk^{(p_i)}_{sig}$
	\State $ACL[pk_{sig}] \gets A[p_i]$
\EndFor
\State $m \gets (addr_P, ACL)$
\State \textbf{send} signed tx(m) to the network\\
\Procedure{StoreIdentity}{$addr_P, ACL$}
	\State $L[addr_P] \gets ACL$
\EndProcedure
\end{algorithmic}
\end{algorithm}

\begin{algorithm}
\caption{Permissions check against the blockchain}\label{alg:check_policy}
\begin{algorithmic}
\Require $pk^{(p_i)}_{sig}$the requesting party signature, $addr_P$ the shared identity's address, $q$ -- a predicate verifying if $p_i$ has sufficient access rights.
\Ensure $s \in \{0,1\}$.
\Procedure{CheckPermission}{$pk^{(p_i)}_{sig}, addr_P, q$}
	\State $s\gets 0$
	\If{$L[addr_P] \neq \emptyset$}
		\State $ACL = L[addr_P]$
		\If{$q(ACL, pk^{(p_i)}_{sig})$}
			\State $s\gets 1$
		\EndIf
	\EndIf
	\State \textbf{return} $s$
\EndProcedure
\end{algorithmic}
\end{algorithm}

\subsubsection{Store and Load}

Storing and loading data for direct access via the DHT are shown in Protocol \ref{alg:data}. For storing data, write permissions are examined with the given $q_{store}$ predicate. The storing party can provide a custom predicate for verifying who can read the data. This is the underlying process that is abstracted away using the $DHT$ singleton object in the scripting language.

\begin{algorithm}
\caption{Storing or Loading Data}\label{alg:data}
\begin{algorithmic}
\Require $pk^{(p_i)}_{sig}$, $addr_P$, $x$ (data), $q_{read}^{(x)}$ -- a predicate for verifying future read access.
\Ensure if successful, returns $a_x$ -- the pointer to the data (predicate), or $\emptyset$ o.w.
\Procedure{Store}{$pk^{(p_i)}_{sig},addr_P,x,q_{read}^{(x)}$}
	\If {$CheckPermission(pk^{(p_i)}_{sig}, addr_P, q_{store} ) = \texttt{True}}$
		\State $a_{x} = \mathcal{H}(addr_P \parallel x)$
		\State $L[a_{x}]\gets q_{read}^{(x)}$
		\State $DHT[a_{x}] \gets x$
		\State \textbf{return} $a_{x}$
	\EndIf
	\State \textbf{return} $\emptyset$
\EndProcedure

\Require $pk^{(p_i)}_{sig}$, $addr_P$, $a_x$ -- the address of the data (predicate)
\Ensure if successful, returns the data $x$, or $\emptyset$ o.w.
\Procedure{Load}{$pk^{(p_i)}_{sig},addr_p, a_x$}
	\State $q_{read}^{(x)} \gets L[a_{x}]$
	\If {$CheckPermission(pk^{(p_i)}_{sig}, addr_P, q_{read}^{(x)} ) = \texttt{True}}$
		\State \textbf{return} $DHT[a_{x}]$
	\EndIf
	\State \textbf{return} $\emptyset$
\EndProcedure
\end{algorithmic}
\end{algorithm}

\subsubsection{Share and Compute}

Share and compute, illustrated in Protocol \ref{alg:share}, are the MPC equivalent of store and load protocols, since they enable processing. Internally, they store and load shares from the DHT and allow working with references to the data while keeping the data secure.

\begin{algorithm}
\caption{Secure computation and secret sharing protocols}\label{alg:share}
\begin{algorithmic}
\Require $pk^{(p_i)}_{sig}$, $addr_P$, $x$ (data), $x_{ref}$ -- reference for computation, $q_{compute}^{(x)}$ -- predicate verifying computation rights.
\Ensure if successful, returns pointer to $x_{ref}$ for future computation, or $\emptyset$ o.w.
\Procedure{Share}{$pk^{p_i}_{sig}, addr_P, x, x_{ref}, q_{compute}^{(x)}, n, t$}
	\State $[x]_p \gets VSS(n, t)$
	\State $peers \gets$ sample $n$ peers
	\For{$peer \in peers$}
		\State \textbf{send} $[x]_p^{(peer)}$ to $peer$ on a secure channel
	\EndFor
	\State \textbf{return} $Store(pk^{(p_i)}_{sig},addr_P,x_{ref},q_{compute}^{(x)})$
\EndProcedure

\Require $pk^{(p_i)}_{sig}$, $addr_P$, $a_{x_{ref}}$ -- reference data address, $f$ -- unsecure code to be rewritten as a secure protocol.
\Ensure if successful, returns $f(x)$ without revealing  $x$, or $\emptyset$ o.w.
\Procedure{Compute}{$pk^{p_i}_{sig}, addr_P, a_{x_{ref}}, f$}
	\State $x_{ref} \gets Load(pk^{(p_i)}_{sig},addr_P, a_{x_{ref}})$
	\If {$x_{ref} \neq \emptyset$}
		\State $f_s \gets$ generate secure computation protocol from $f$
		\State \textbf{return} $f_s(x_{ref})$
	\EndIf
	\State \textbf{return} $\emptyset$
\EndProcedure
\end{algorithmic}
\end{algorithm}



\section{Incentives}

Since Enigma is not a cryptocurrency or a blockchain, the incentive scheme is based on fees rather than mining rewards, where nodes are compensated for providing computational resources. Full nodes are required to provide a security deposit, making malicious behaviour punishable.

\subsection{Security Deposits}

A possible attack on MPC protocols takes advantage of the lack of guaranteed fairness in the protocol. Under certain conditions, a malicious party can learn the output and abort the protocol before other parties learn the output as well. While this attack, when carried out by a majority, cannot be prevented, it can be penalized. Using Bitcoin security deposits for punishing malicious nodes in MPC has been investigated by several scholars recently \cite{deposits1, deposits2}. We use a similar model, and extend it to penalize other malicious behaviors such as breaking correctness, which is validated by the SPDZ protocol (see section \ref{sec:comp_correct}).

To participate in the network, store data, perform computations and receive fees, every full-node must first submit a security deposit to a private contract. After each computation is completed, a private contract verifies correctness and fairness were maintained. If a node is found to lie about their outcome or aborts the computation prematurely, it loses the deposit which is split between the other honest nodes. The computation is continued without the malicious node (e.g., by setting its share of the data to 0).

\subsection{Computation Fees}

Every request in the network for storage, data retrieval, or computation has a fixed price, similar to the concept of ‘Gas’ in Ethereum. Unlike Ethereum where every computation is run by every node, in Enigma different nodes execute different parts of each computation and need to be compensated according to their contribution, which is measured in rounds. Recall that every function is reduced to a circuit of addition and multiplication gates, each of which takes one or more rounds. A node participating in a computation is paid the weighted sum of the number of rounds it contributed to and the operations it performed (addition, multiplication).


Since the platform is turing-complete the exact cost of a request cannot always be pre-calculated. Therefore, once the computation is finalized, the cost of each request is deducted from an account balance each node maintains. A request will not go through unless the account balance is over a minimum threshold.

\subsection{Storage Fees}

Fees for data storage are market based and time limited. The hosting contract is automatically renewed using the owner's account balance. If the balance is too low, access to the data will be restricted and unless additional funds are deposited, the data will be deleted within a certain amount of time.

\section{Applications}

\subsection{Data Marketplace}

Direct consumer to business marketplace for data. With guaranteed privacy, autonomous control and increased security, consumers will sell access to their data. For example, a pharmaceutical company looking for patients for clinical trials can scan genomic databases for candidates. The marketplace would eliminate tremendous amounts of friction, lower costs for customer acquisition and offer a new income stream for consumers.

\subsection{Secure Backend}

Many companies today store large amounts of customer data. They use the data to provide personalized services, match individual preferences, target ads and offers, etc. With Enigma, companies can use the data for the same purposes they do today, without actually storing or processing the data on their servers, removing security risks and assuring the privacy of their customers. 

\subsection{Internal Compartmentalization}

Large organizations can use Enigma to protect their data and trade secrets from corporate espionage and rogue employees. Employees can still use and analyze data for the benefit of the organization, but won`t be able to steal any data. Productivity inside organizations would be improved since more people can have access to more data, and costs on security would be lower.

\subsection{N-Factor Authentication}

Voice, face and fingerprint recognition stored and computed on Enigma. Only the user ever has access to these data. Policies for when and if additional keys are required can be set inside a private contract, unexposed to any potential attacker.

\subsection{Identity}

Authenticating and securely storing identities in a fully anonymous, yet provably correct, fashion is trivial on Enigma and requires as little as several lines of code. The process is simple -- a user secret-shares her personal information required for authentication. When the user logs in, an authenticating private contract is executed, validating the user and linking her real identity with a public pseudo-identity. The process is completely trust-less and privacy-preserving.

\subsection{IoT}

Store, manage and use (the highly sensitive) data collected by IoT devices in a decentralized, trust-less cloud.


\subsection{Distributed Personal Data Stores}

Store and share data with third parties while maintaining control and ownership. Set specific policies for each service with private contracts. Identity is truly protected since the decision to share data is always reversible - services have no access to raw data, all they can do is run secure computations on it. 

\subsection{Crypto Bank}

Run a full-service crypto bank without exposing private internal details. Users can take loans, deposit cryptocurrencies or buy investment products with the autonomous control of the blockchain, without publicly revealing their financial situation.

\subsection{Blind E-Voting}

Votes on anything, from political elections to company board meetings, without exposing anything besides the final outcome. Not only is the privacy of each voter is maintained, even the actual vote-count can remain private. For example, if the elections require any kind of majority vote, but no details about the distribution, a unanimous decision would be indistinguishable from one decided by a single vote.

\subsection{Bitcoin Wallet}

\begin{enumerate}
    \item Decentralized private key generation -- Multiple Enigma nodes locally create a segment of the key, whereas the full key is only ever assembled by the user. No trail of evidence is left anywhere.
    
    \item Decentralized transaction signing -- Transactions signed without ever exposing the private key or leaving a trail.
    
    \item Decentralized controls - Set spending limits, multi-sig, CHECKLOCKTIMEVERIFY like controls, and more with a private script. Lock time, limits or number of required signatures are completely invisible to a potential attacker.
    
\end{enumerate}




\end{document}